\documentclass[12pt,preprint]{aastex}

\shorttitle{Cosmic rays from active galactic nuclei}
\shortauthors{E.G. Berezhko }

\begin{document}
\title{Cosmic rays from active galactic nuclei}

\author{E.G.Berezhko\altaffilmark{1}}
\altaffiltext{1}{Yu.G. Shafer Institute of Cosmophysical Research and Aeronomy,
                     31 Lenin Ave., 677980 Yakutsk, Russia}

\email{berezhko@ikfia.ysn.ru}

\begin{abstract} 
Cosmic ray (CR) acceleration at the shock created by the expanding cocoons
around active galactic nuclei (AGNs) is studied. It is shown that 
above the energy $10^{18}$~eV the overall 
energy spectrum of CRs, produced during the AGN evolution and 
released in the intergalactic space,
 has the form $N\propto \epsilon^{-\gamma}$,
with $\gamma\approx 2.6$, which extends up to $\epsilon_\mathrm{max}\sim 10^{20}$~eV.
It is concluded that cocoons shocks have to be considered as a main source of
extragalactic CRs, which together with Galactic supernova remnants provide
the observed CR spectrum.
\end{abstract}


%

\section{Introduction}
The overall origin of cosmic rays (CRs) is still 
an unresolved problem in astrophysics.
The understanding of CR origin requires the determination of 
the astrophysical objects,
that are the CR sources, and of the appropriate acceleration processes, 
that form the CR spectrum in these objects.
In this regard, during the last several years considerable progress 
has been achieved in this field, both
experimentally and theoretically.
Recently the steepening of the CR spectrum above $3\times 10^{19}$~eV
was established
in the HiRes \citep{hires07} and Auger \citep{augerJ07} experiments.
It presumably corresponds to the so-called
Greizen-Zatsepin-Kuzmin (GZK) cutoff, caused by CR energy
losses in their interactions with the microwave background radiation.
This is evidence that the highest energy part of the CR spectrum
is of extragalactic origin.
It was also recently demonstrated \citep{bv07}
that the CRs with energies
up to $\epsilon \sim 10^{17}$~eV can be  
produced in supernova remnants (SNRs).

The main reason why SNRs are usually considered as
the CR source is a simple argument about the
energy required to sustain the Galactic CR population against
loss by escape, nuclear interactions and ionization energy loss. 
Supernovae have enough power to drive the CR acceleration. 
The high velocity ejecta 
produced in the supernova 
explosion interacts with the ambient medium to produce a strong blast wave, 
which may accelerate a
small suprathermal fraction of the ambient plasma to high energies.
The only theory of particle acceleration which is sufficiently well
developed and specific to allow quantitative model calculations is diffusive
acceleration \citep{krym77,bell78} applied to the strong outer shock associated
with SNRs \citep[e.g. see][for review]{bek88,ber05,ber07}. 
Considerable efforts have been
made during the last years to empirically confirm the theoretical expectation
that the main part of CRs indeed originates in SNRs. Theoretically, progress
in the solution of this problem has been due to the development of 
a kinetic nonlinear theory of diffusive shock acceleration
\citep{byk96}. This theory includes
all the most relevant physical factors, essential for the SNR evolution
and CR acceleration in a SNR and it is able to make quantitative predictions of the expected
properties of CRs produced in young SNRs and their nonthermal
radiation. 
Applied to individual young SNRs
\citep[see][for reviews]{ber05, ber07} this theory has successfully
explained many observed SNR properties. Making use of the observed synchrotron
emission spectrum from radio to X-ray frequencies 
it permits the determination of the degree of
magnetic field amplification,
a process advocated earlier from plasma simulations \citep{lb00,belll01}, 
which leads to the concentration of the
highest-energy electrons in a very thin shell just behind the shock. 
Such observed filamentary structures have been observed
in hard X-rays \citep{vl03,long03,bamba03,par06} 
and are nowadays used as a second
independent method to infer the magnitude of the amplified field
\citep[see][for interpretation]{bkv03}.

It has been demonstrated, that magnetic field amplification leads to considerable
increase of the maximal energy of CRs accelerated in SNRs and
that the observed CR energy spectrum can be well 
represented by two components \citep{bv07}. 
The first one, $J^{g}(\epsilon)$, 
dominant up to $10^{17}$~eV, consists of CRs
produced in Galactic SNRs, whereas the second, 
$J^{eg}(\epsilon)$, is produced in extragalactic
sources. This so called dip scenario \citep{aloisio06} 
requires a relatively steep CR spectrum
produced in extragalactic sources, $J_\mathrm{s}^{eg}\propto \epsilon^{-\gamma}$ 
with $\gamma =2.55- 2.75$ at 
energies $\epsilon>10^{18}$~eV. 
It has been demonstrated that peculiarities in the CR spectrum 
-- the so-called "dip" structure at $\epsilon\sim 10^{19}$~eV 
and a GZK cutoff at
$\epsilon>3\times 10^{19}$~eV \citep{aloisio06} --
produced due to CR 
interaction with the
cosmic microwave background radiation on their way from the source to the
observer are very consistent with the experiment.

Below we consider CR acceleration at the shock created by the expanding cocoon
around active galactic nuclei (AGNs).
Compared with previous considerations \citep{norman95} magnetic field
amplification is taken into account, which provides CR acceleration with an
appropriate spectrum up to the energy $~ 10^{20}$~eV. 
Therefore the cocoon
shocks have to be considered as a main source of
extragalactic CRs, which together with Galactic supernova remnants 
presumably provide the observed CR spectrum.

\section{Results and discussion}
Astrophysical objects, which are considered as potential extragalactic sources
of ultra high energy CRs, should fulfill a 
number of requirements.
First, they should have sufficiently high energy output, not less than
$L_\mathrm{p}\sim 2\times 10^{45}-3.5\times 10^{46}$~erg~Mpc$^{-3}$yr$^{-1}$ 
depending on the form
of CR spectrum that they produce \citep{berez06}.

Given the energy
requirements, AGNs \citep[e.g.][]{berez06} and
gamma-ray bursts (GRBs) \citep{milgrom95,waxman95,vietri95} are considered as
potential extragalactic sources of the ultra-high-energy
CRs. However, as \citet{berez06} argued, 
the energy output of GRBs has a
serious problem if they are considered as 
the main source of the extragalactic CRs.

The second requirement to the
extragalactic CR sources is their ability to produce
a power-law CR spectrum up to the maximal energy, which for protons 
should be at least as large as
$\epsilon_\mathrm{max}=10^{20}$~eV, which is well above the GZK cutoff.
Particle acceleration at relativistic shocks in AGNs 
\citep{achterberg00,rb93} and 
GRBs \citep{milgrom95,waxman95,vietri95} are discussed 
as appropriate acceleration processes.
However, simulations of CR
shock acceleration \citep{niemiec06} have demonstrated 
a low efficiency in the case of relativistic shocks.

The relativistic jet in AGNs is surrounded by the hot cocoon, which expands
into the surrounding intergalactic medium
with supersonic speed \citep[e.g.][]{ferrari}.
Since the essential fraction of the jet 
energy is transformed into the internal
energy of the background medium due to the outermost nonrelativistic shock,
driven by the cocoon overpressure,
it is natural to suppose that a  good part of this energy is represented
by effectively accelerated CRs, like what takes place in SNRs.
Since CR acceleration by nonrelativistic shocks 
has been very well studied
one can obtain reliable estimates of the expected spectrum 
of CRs.
First of all 
the maximal energy of CRs, accelerated at the expanded shock of
size $R(t)$ and speed $V=dR/dt$, is determined by the expression \citep{ber96}
\begin{equation}
\epsilon_\mathrm{max}\approx ZeBRV/c,
\end{equation}
where $B$ is the upstream magnetic field, $c$ is the speed of light, $e$ is the
proton charge and $Z$ is the charge number of CR nuclei.
This expression consistently takes into account the influence of all the most
essential physical factors restricted the maximal CR energy, which are the adiabatic
cooling of CRs in the downstream region and the shock deceleration followed by the
CR escape from the shock vicinity into outer space \citep{ber96}.

Magnetic field near the shock front, as was established for all young SNRs
\citep{ber05,ber07}, is strongly amplified nonlinearly by CRs up to the level
\begin{equation}
B^2/(8\pi)\approx 3\times 10^{-3}\rho V^2,
\end{equation}
which is presumably also appropriate for extragalactic shocks.
Here $\rho$ is the external gas density.
Note that the relation $B\propto V$ following from Eq.(2) is indeed theoretically
expected for the shock speeds $V<10^4$~km s${-1}$ \citep{pell06}.
Due to the nonresonantly amplified magnetic field, as was originally shown by
\citet{bell04}\citep[see also][]{pell06}, at higher shock speed $V>10^4$~km s${-1}$
the expected relation is $B\propto V^{3/2}$, which leads to larger magnetic field
strength compared with the value given by Eq.(2). Due to this fact CR maximal
energy $\epsilon_\mathrm{max}$
can be a few times larger compared with our estimate below. However 
this does not play a critical role because even our estimate gives 
$\epsilon_\mathrm{max}\approx 10^{20}$~eV, which is well above GZK cutoff.

The outer cocoon shock expands
with the speed \citep[e.g.][]{ferrari}
\begin{equation}
V\approx [L_\mathrm{j}/(\rho V_\mathrm{h})]^{1/4}t^{-1/2},
\end{equation}
where $L_\mathrm{j}$ is the mechanical luminosity of the jet and 
$V_\mathrm{h}$ is the hot 
spot or the jet head speed. 
Substituting the magnetic field value following 
from Eq.(2), the shock speed (3) and
the shock size $R=2Vt$ into the equation (1), we have
\[
\epsilon_\mathrm{max}\approx 10^{20} Z \left(
\frac{L_\mathrm{j}}{10^{46}~\mbox{erg/s}}\right)^{3/4}
\left(
\frac{N_\mathrm{g}}{10^{-4}~\mbox{cm}^{-3}}\right)^{-1/4}
\]
\begin{equation}
\hspace{0.5cm}\times
\left(
\frac{V_\mathrm{h}}{10^{10}~\mbox{cm/c}}\right)^{-1/4}
\left(
\frac{l_\mathrm{j}}{1~\mbox{kpc}}\right)^{-1/2}~\mbox{eV},
\end{equation}
where $l_\mathrm{j}=V_\mathrm{h}t$ is the jet length. 
According to this expression,
during the evolutionary epochs corresponding to the jet
length from $l_\mathrm{j}=1$~kpc to $l_\mathrm{j}=10$~Mpc
the cocoon shock produces a power-law proton spectrum which extends up to the
maximal energy from $\epsilon_\mathrm{max}=10^{18}$~eV to
$\epsilon_\mathrm{max}=10^{20}$~eV.
Thus we conclude, that AGNs fulfill also the second requirement for
extragalactic CR sources:
protons, the main kind of ions in the space plasma,
are effectively
accelerated at least up to the energy $\epsilon_\mathrm{max}=10^{20}$~eV. 
Heavier ions are expected to be accelerated to $Z$ times higher maximal energy.
However, since the GZK cutoff energy of nuclei
$\epsilon_\mathrm{GZK}^A<\epsilon_\mathrm{GZK}$ for
the atomic number
$A>1$ is lower then that for protons $\epsilon_\mathrm{GZK}\approx 5\times 10^{19}$~eV
\citep{berez06}, the predicted dependence
$\epsilon_\mathrm{max}\propto Z$ is difficult to establish experimentally.

The most realistic possibility of finding 
the signature of the acceleration process
of ultra-high-energy CRs is the
experimental study of CR composition at energies 
$\epsilon>10^{17}$~eV. If extragalactic CRs are indeed produced by large-scale
nonrelativistic shocks their composition will have the following peculiarity.
Inside and near the source, CR composition corresponds to the composition of the
intergalactic gas with substantial enrichment by heavy elements, like with Galactic
CR versus interstellar medium compositions \citep[e.g.][]{byk96}.
If compositions of intergalactic and interstellar plasmas are not very
different, then the expected source CR composition is characterised by the mean
atomic number $<A>\approx 5$, like Galactic CR components.
Compared with this source composition, the observed composition is significantly
distorted at almost all energies due to propagation effects in the intergalactic
medium and in the Galactic wind. Only at the energy 
$\epsilon_*\approx 2\times 10^{18}$~eV are these compositions roughly the same
because it is just below the lowest GZK cutoff energy 
$\epsilon_\mathrm{GZK}^\mathrm{He}\approx 4\times 10^{18}$~eV 
and the modification of the
proton spectrum at this energy is still not very large \citep{berez06}.
Since CR diffusion mobility for given energy $\epsilon$ is lower for larger $Z$,
heavier CRs undergo stronger modification (depression) during their propagation
from the sources to the observer. Due to this effect composition of
extragalactic CRs is expected to become progressively lighter with the decrease
of their energy at $\epsilon<\epsilon_*$. At higher energies,
$\epsilon>\epsilon_*$, due to the dependence of the GZK cutoff energy
$\epsilon_\mathrm{GZK}\propto A$ on the nuclear mass number $A$, CRs become
progressively heavier so that the peak value of $<A>$ is expected at the
beginning of the GZK cutoff for iron, which is at $\epsilon\approx 10^{19}$~eV
\citep{berez06}. At such a high energies only protons and iron nuclei
survive and the iron-to-proton ratio has a maximum value at the energy 
$\epsilon\approx 10^{19}$~eV, which provides a peak of CR mean mass very
similar to the peak at $\epsilon\approx 10^{17}$~eV. To make a quantitative
prediction of the expected ultra-high-energy CR composition, one has to perform
a detailed consideration, which will be done in a subsequent paper.

Note that the expression (4)
is based on the amplified magnetic field value (2) and therefore
does not depend on the assumed value of intergalactic magnetic field
in contrast to the earlier estimate 
of $\epsilon_\mathrm{max}$ \citep{norman95}.
Note also that as is seen from the Eq.(4), maximal CR energy only slightly
depends on external gas density and the jet head speed.
The values of mechanical luminosity $L_\mathrm{j}$ and the jet length
$l_\mathrm{j}$, 
which determines the maximal
CR energy, can be directly measured in the experiment.

The third requirement to CR sources is related to the form of the CR spectrum.
The form of the resulting CR spectrum produced during the whole evolution of the
expanding shock is determined by three physical factors:
(1) nonlinear shock modification due to the CR back-reaction, (2) adiabatic energy
losses in the downstream region, and (3) diffusive CR escape from the shock vicinity into
outer space. The existence of the CR escape phenomenon
makes it possible to estimate the shape of the spectrum of 
the most energetic CRs produced during the source evolution.
As was demonstrated analytically
\citep{bek88,ber96} and confirmed numerically \citep{byk96}, 
since the maximal energy $\epsilon_\mathrm{max}(t)$ of CRs accelerated by the expanding
shock during its most active phase decreases with time $t$ due to shock
deceleration, the most energetic
particles with energy
$\epsilon\sim \epsilon_\mathrm{max}(t)$ undergo diffusive escape from the shock 
vicinity into the
surrounding space. If such behavior happens during the time period from $t_1$ to
$t_2$, then at any given epoch $t_1<t<t_2$ all accelerated particles with
energies $\epsilon >\epsilon_\mathrm{max}(t)$ have already escaped and inside the source
there are only particles with $\epsilon \le\epsilon_\mathrm{max}(t)$.

Note that the particle escape phenomenon, has two important aspects.
First, the energy of escaped CRs determined for any evolutionary epoch by Eq.(1)
does not depend any more on the
adiabatic cooling inside the parent expanding source, 
like what happens with CRs of
lower energies $\epsilon <\epsilon_\mathrm{max}(t)$.
Second, the spectrum of the escaped CRs can be essentially different compared with
the canonical spectrum $N\propto \epsilon^{-2}$, produced by 
strong nonrelativistic shock.
To estimate the spectrum of escaped CRs one can use 
the simple relation \citep{bek88}
\begin{equation}
N(\epsilon)\epsilon d\epsilon
\propto \rho V^2R^{2 -\beta}dR,
\end{equation}
which determines at any given phase $t$
the overall number of CRs $N(\epsilon)$
with highest energy $\epsilon\sim \epsilon_\mathrm{max}(t)$. Due to the hard
self-consistent spectrum of CRs, produced by the strong shock,
the main contribution to the CR energy content is given by the particles
with highest energy $\epsilon\sim \epsilon_\mathrm{max}(t)$ \citep{byk96}. 
Therefore their energy content
scales as the shock ram pressure $\rho V^2$ times the shock volume, 
as it is given by Eq.(5).
Factor $R^{-\beta}$ describes the progressive
slow decrease of CR acceleration efficiency due to
the shock weakening.
For the shock expansion law $R\propto t^{-\nu}$ relation (5) gives
\begin{equation}
N(\epsilon)
\propto \epsilon^{-\gamma}~~~\mbox{with}~~~\gamma=1+(2-\beta)/(2/\nu-3).
\end{equation}
In our case $\nu=1/2$ this gives $\gamma=3-\beta$.
For the constant acceleration efficiency 
we have $\beta=0$ which gives $\gamma=3$.
However, the shock deceleration is accompanied by the decrease of the acceleration
efficiency. This effect is described by the amount of shock modification, which is
characterized by the shock compression ratio, which for the case of a strong shock
depends on the shock speed as
$\sigma \propto V^{3/8}$ \citep{byk96}. Since in our case $V\propto 1/R$ we 
have $\beta=3/8$ that
gives $\gamma\approx 2.6$. Such a spectrum of extragalactic CRs at energies
$\epsilon >10^{18}$~eV corresponds very well to the experiment \citep{berez06}.

Particles with energies $\epsilon \le\epsilon_\mathrm{max}(t)$ at any given epoch
are contained inside the source and have a spectrum close to
$N\propto \epsilon^{-2}$. At a very late epoch when the outer shock becomes weak
these particles will leave the source. Note however that this is already 
not important
for the Galaxy, because the contribution of extragalactic CRs is expected 
to be low at
energies $\epsilon <10^{18}$~eV \citep{berez06}.

Note that based on the data collected at the Auger experiment, a
correlation between the arrival directions of CRs with energy above $6\times
10^{19}$~eV and the position of nearby AGNs has been found \citep{augercol07},
which strongly supports AGNs as prime candidates for the source of 
ultra-high-energy CRs.

\section{Summary}
Magnetic field amplification due to shock-accelerated CRs
provides acceleration of CRs up to the energy
$\epsilon_\mathrm{max}\sim 10^{20}$~eV at
the outer nonrelativistic shock created by the expanding cocoon around AGNs . 

The  expected CR spectrum formed due their escape from the expanding shock
has a form $N\propto \epsilon^{-\gamma}$ with $\gamma\approx 2.6$
at energies
$\epsilon> 10^{18}$~eV, which
fulfills quite well the requirements for extragalactic CR sources.
Therefore AGNs
have to be considered as a prime
candidate for the sources of extragalactic CRs. 
In such a case the observed CRs consist of two components:
Galactic CRs, produced in SNRs, and extragalactic CRs from AGNs.

The expected composition of ultra-high-energy CRs is characterised by two peaks
in the energy dependence of the mean CR atomic number $<A(\epsilon)>$.
The first one at the energy $\epsilon\approx10^{17}$~eV corresponds to the very
end of the Galactic CR component \citep{bv07}, whereas the second, at the energy 
$\epsilon\approx10^{19}$~eV, is expected at the beginning of the GZK cutoff.
The strong energy dependence of CR composition within the energy interval
from $10^{17}$ to $10^{18}$~eV is expected as a 
signature of the transition from Galactic to
extragalactic CRs, whereas the detection of heavy CR composition at
$\epsilon\approx10^{19}$~eV has to be considered as the signature of CR
production by nonrelativistic shocks.

\acknowledgements
The work was supported by the 
Presidium of the RAS (program 16) and by
the Russian Foundation for Basic Research (grants 06-02-96008 and 07-02-00221).


\begin{thebibliography}{}

\bibitem[Abraham et al.(2007)]{augercol07}
Abraham, J. (Pier Auger Collaboration) 2007, Science, 318, 938 

\bibitem[Achterberg et al.(2000)]{achterberg00}
Achterberg, A., Gallant, Y.A., Kirk, J.G., \& Guthmann, A.W. 2000,
MNRAS, 328, 393  

\bibitem[Aloisio et al.(2007)]{aloisio06}
Aloisio, R., et al. 2007, Astropart. Phys., 27, 76 

\bibitem[Bamba et al.(2003)]{bamba03}
Bamba, A., Yamazaki, R., Ueno, M., \& Koyama, K. 2003, ApJ, 589, 827

\bibitem[Bell(1978)]{bell78}
Bell, A.R. 1978, MNRAS, 182, 147 

\bibitem[Bell \& Lucek(2001)]{belll01}
Bell, A. R., \& Lucek, S.G. 2001, MNRAS, 321, 433

\bibitem[Bell(2004)]{bell04}
Bell, A.R. 2004, MNRAS, 353, 550

\bibitem[Berezhko(1996)]{ber96}
Berezhko, E.G. 1996, Astropart. Phys., 5, 367 

\bibitem[Berezhko(2005)]{ber05}
Berezhko, E.G. 2005, Adv. Space Res., 35, 1031 

\bibitem[Berezhko(2008)]{ber07}
Berezhko, E.G. 2008, Adv. Space Res., 41, 429 

\bibitem[Berezhko \& Krymsky(1988)]{bek88}
Berezhko,  E.G., \& Krymsky, G.F. 1988, Soviet Phys.-Uspekhi, 12, 155 

\bibitem[Berezhko et al.(1996)]{byk96}
Berezhko, E.G., Elshin,  V.K., \& Ksenofontov, L.T. 1996, JETPh, 82, 1 

\bibitem[Berezhko et al.(2003)]{bkv03}
Berezhko, E. G., Ksenofontov, L. T., \& V\"olk, H. J. 2003, A\&A,
412, L11

\bibitem[Berezhko \& V\"olk(2007)]{bv07}
Berezhko, E.G. \&  V\"olk, H.J. 2007, Astrophys. J., 661, L175 

\bibitem[Berezinsky et al.(2006)]{berez06}
Berezinsky, V.S., Gazizov, A., \& Grigorieva, S. 2006,
Phys. Rev. D, 74, 043005 

\bibitem[Bergman et al.(2007)]{hires07}
Bergman, D.R. et al. 2007 Nuclear Phys. B (Proc. Supl.), 165, 19

\bibitem[Ferrari(1998)]{ferrari}
Ferrari, A. 1998, Annu. Rev. Astron. Astrophys., 36, 539 

\bibitem[Krymsky(1977)]{krym77}
Krymsky, G.F. 1977, Soviet Phys. Dokl., 23, 327

\bibitem[Long et al.(2003)]{long03}
Long, K.S., Reynolds, S.P., Raymond, J.C., et al. 2003, ApJ, 586, 1162

\bibitem[Lucek \& Bell(2000)]{lb00}
Lucek, S.G., \& Bell, A.R. 2000, MNRAS, 314, 65

\bibitem[Milgrom \& Usov(1995)]{milgrom95}
Milgrom, M. \& Usov, V. 1995, ApJ, 449, L37 

\bibitem[Niemiec et al.(2006)]{niemiec06}
J. Niemiec, J., Ostrowski, M., \& Pohl, M. 2006, ApJ, 650, 1020 

\bibitem[Norman et al.(1995)]{norman95}
Norman, C.A., Melrose, D.B. \& Achterberg, A. 1995, ApJ, 454, 60

\bibitem[Parizot et al.(2006)]{par06}
Parizot, E., Marcowith, A., Ballet, J., \& Gallant, Y.A. 2006, A\&A, 453, 387

\bibitem[Pelletier et al.(2006)]{pell06}
Pelletier, G. Lemoine, M., \& Marcovith, A. 2006, A\&A, 453, 181

\bibitem[Rachen \& Biermann(1993)]{rb93}
Rachen, J.P., \& Biermann, P. 1993, A\&A, 272, 161 

\bibitem[Vietri(1995)]{vietri95}
Vietri, M. 1995, ApJ., 453, 883 

\bibitem[Vink \& Laming(2003)]{vl03}
Vink, J., \& Laming, J.M. 2003, ApJ, 548, 758

\bibitem[Waxman(1995)]{waxman95}
Waxman, E. 1995, Phys. Rev. Let., 75, 386 

\bibitem[Yamamoto et al.(2007)]{augerJ07}
Yamamoto, T. et al. arXiv:0707.2638v3[astro-ph](2007)

\end{thebibliography}
\end{document}